\documentstyle[aps,epsfig,12pt]{revtex}
\def\bege{\begin{equation}}
\def\ende{\end{equation}}

\draft
\begin{document}
\tightenlines
\title{\bf Nonperturbative Vacuum Effect in the Quantum Field Theory of 
Meson Mixing } \author{  Chueng-Ryong Ji and Yuriy Mishchenko }
\address{ Department of Physics, North Carolina State University \\
          Raleigh, North Carolina 27695-8202 USA \\ }
\maketitle

\begin{abstract}
Replacing the perturbative vacuum
by the nonperturbative vacuum, we extend a recent development of
a quantum field theoretic framework for scalar and
pseudoscalar meson mixing. The unitary inequivalence of the Fock
space of base (unmixed) eigenstates and the physical mixed eigenstates
is investigated and the flavor vacuum state structure is explicitly found. 
This is exploited to develop formulas for two flavor boson oscillations in
systems of arbitrary boson occupation number. We apply these formulas to
analyze the mixing of $\eta$ with ${\eta}'$ and comment on the other 
meson-mixing systems. In addition, we consider the mixing of boson coherent
states, which may have future applications in the
construction of meson lasers.
\end{abstract}
\newpage
\baselineskip=20pt
\setcounter{section}{0}
\setcounter{equation}{0}
\setcounter{figure}{0}
\renewcommand{\theequation}{\mbox{1.\arabic{equation}}}
\renewcommand{\thefigure}{\mbox{1.\arabic{figure}}}
\section{Introduction}
The study of mixing transformations plays an important part 
in particle physics phenomenology.{\cite{1}} The Standard Model 
incorporates the mixing of fermion
fields through the Kobayashi-Maskawa{\cite{2}} mixing of 3 quark flavors,
a generalization of the original Cabibbo{\cite{3}} mixing matrix between
the $d$ and $s$ quarks. In addition, neutrino mixing
and oscillations are the likely resolution of the famous solar neutrino
puzzle{\cite{4}}. In the boson sector, the mixing of $K^0$ with
$\overline{K^0}$ via weak currents provided the first evidence of $CP$
violation{\cite{5}} and the $B^0\bar{B}^0$ mixing plays an important role
in determining the precise profile of a CKM {\cite{2,3}}
unitary triangle \cite{ref1} in Wolfenstein parameter space \cite{ref2}. 
The ${\eta}$ ${\eta}'$ mixing in the $SU(3)$
flavor group also provides a unique opportunity for testing QCD and the
constituent quark model. Furthermore, the particle mixing relations for
both the fermion and boson case are believed to be related to the
condensate structure of the vacuum. The non-trivial nature of the vacuum
is expected to hold the answer to many of the most salient questions
regarding confinement and the symmetry breaking mechanism.

The importance of the fermion mixing transformations has recently prompted
a fundamental examination of them from a quantum field theoretic
perspective{\cite{6}}. A similar analysis in the bosonic sector has also
been undertaken{\cite{7}}. However, more recent analysis{\cite{8}} on the
fermion mixing indicated that the previous result{\cite{6}} based on the
perturbative vacuum is only the approximation with respect to the exact one
based on the nonperturbative (flavor) vacuum. In this work, we show that
the same is true for the bosonic 
sector. Upon the completion of our work, we notice that the same conclusion
was also drawn in a recent literature \cite{Blasson3}. In our work, 
however, the orthogonality between mass and flavor vacua is shown in a 
straightforward algebraic method rather than solving a differential 
equation for the inner
product of two vacua as presented in \cite{Blasson3}.
As evidenced in the previous literatures \cite{6,8,Blasson3,25} the
method of using a differential equation to prove the unitary 
inequivalence between the two Fock spaces has been known for some time 
and our algebraic method is a new development in this respect. Moreover, 
we analyze the structure of the nonperturbative flavor vacuum in a great 
detail contrasting to the fermion case. The details of flavor vacuum, 
its perturbative expansion in the mixing angle and also some clarifying 
remarks on the Green
function method and the arbitrary mass parametrization are 
summarized in the accompanying Appendices.

We begin in Section II with investigation of the vacuum structure
using the relation between the base eigenstate
and the physical mixed-eigenstate fields.
We derive the representation for Pontecorvo mixing transformation
for boson case and explicitly calculate 
the flavor vacuum state structure in the quantum field theory (QFT). 
We then investigate the unitary inequivalence of the
two Fock spaces - one is the space of mass eigenstates and the other 
is the space of flavor eigenstates.
In Section III, the ladder operators
are constructed in the mixed basis. These are used to derive time
dependent oscillation formulas for $1$-boson states, $n$-boson states,
and boson coherent states.
Consequences from the replacement of the perturbative vacuum by the
exact nonperturbative (flavor) vacuum are demonstrated.
Section IV is devoted to study specific cases in our formalism, such as
the ${\eta}$ ${\eta}'$ system. We show the numerical differences between
the two results: one from the perturbative vacuum and the other 
from the nonperturbative vacuum. Conclusions and discussions follow 
in Section V. In Appendix A, we present a derivation of an explicit 
expression for the flavor vacuum operating the ladder operators of 
particle and antiparticle to the vacuum of mass eigenstates. 
In Appendix B, we discuss the region of validity for
a perturbative expansion of the flavor vacuum. In Appendix C,
we make some clarifying remarks on the Green function method and 
the arbitrary mass parametrization discussed in 
recent literatures\cite{8,Blasson3,ref3}. 
\setcounter{equation}{0}
\setcounter{figure}{0}
\renewcommand{\theequation}{\mbox{2.\arabic{equation}}}
\renewcommand{\thefigure}{\mbox{2.\arabic{figure}}}
\section{The Mixing Relation and Vacuum Structure}
We start our analysis by considering the Pontecorvo mixing relationship
\cite{Pontecorvo2_5} for two fields:
\begin{eqnarray}\label{transf}
\phi_{\alpha}&=&\cos\theta\varphi_{1}+\sin\theta\varphi_{2}\cr
\phi_{\beta}&=&-\sin\theta\varphi_{1}+\cos\theta\varphi_{2},
\end{eqnarray}
where $\varphi_{1,2}$ are the free fields with definite masses $m_{1,2}$
and $\phi_{\alpha,\beta}$ are the interacting fields with 
definite flavors $\alpha$, $\beta$, respectively. The above mentioned 
relationship naturally arises by considering
the mixing problem for the two quantum fields with the lagrangian of the form
\begin{equation}\label{lagr}
L=L_{0,\alpha}+L_{0,\beta}-\lambda (\phi_{\alpha}^{\dag} \phi_{\beta}
       +\phi_{\beta}^{\dag} \phi_{\alpha}),
\end{equation}
where $L_{0,\alpha (\beta)}$ are the free flavor-field lagrangians 
(i.e. $L_{0, \alpha (\beta)}=\frac{1}{2}(\partial \phi_{\alpha (\beta)}^{\dag}
\partial \phi_{\alpha (\beta)}
-m_{\alpha (\beta)}^2\phi_{\alpha (\beta)}^{\dag}\phi_{\alpha (\beta)})$)
and $\lambda$ is the coupling
constant responsible for mixing. It is straightforward to show that the above
lagrangian can be immediately diagonalized by the transformation given by Eq.
(\ref{transf}) with an appropriate choice of mixing angle $\theta$
\cite{8}. The parameters of diagonalized lagrangian can then be  
expressed in terms of 
flavor-field masses ($m_{\alpha},m_{\beta}$) and interaction constant 
($\lambda$); i.e.
\begin{equation}
\tan(2\theta)=\frac{4 \lambda}{m_{\alpha}^2-m_{\beta}^2}
\end{equation}
 and
\begin{equation}
m_{1,2}^{2}=\frac{(m_{\alpha}^2+m_{\beta}^2)\pm
 \sqrt{(m_{\alpha}^2-m_{\beta}^2)^2+16\lambda^2}}{2}.
\end{equation}

The free mass-fields $\varphi_{1,2}$ can be written explicitly as usual:
\begin{equation}\label{field}
\varphi_{i}=\sum_{\vec{k}}\frac{1}{\sqrt{2\epsilon_{i}(k)}}(u_{\vec{k}i}
a_{\vec{k}i}e^{-i k x}+v_{\vec{k}i}b_{\vec{k}i}^{\dag}e^{i k x}),
\end{equation}
where $a_{\vec{k}i}$ and $b_{\vec{k}i}$ are respectively the particle 
and anti-particle ladder operators for the free mass-fields and they satisfy 
the standard equal-time  commutation relationships:
\begin{eqnarray}\label{commutation_m}
[a_{\vec{k}i},a_{\vec{k}'i'}^{\dag}]&=&
\delta_{\vec{k},\vec{k}'}\delta_{i,i'} \cr
[b_{\vec{k}i},b_{\vec{k}'i'}^{\dag}]&=&
\delta_{\vec{k},\vec{k}'}\delta_{i,i'}.
\end{eqnarray}
Here, $k x=k_{0} x_{0}-\vec{k}\cdot\vec{x}$ and 
$\epsilon_{i}(k)=k_{0}(k)=\sqrt{\vec{k}^2+m_{i}^2}$. For the spin-0 case, 
free mass-field amplitudes
$u_{\vec{k}i}$ and $v_{\vec{k}i}$ are just numbers, i.e.
$u_{\vec{k}i}=v_{\vec{k}i}=1$.
The interacting flavor-fields given by Eq.(\ref{transf}) are the solutions of 
the Euler-Lagrange equation for Eq.(\ref{lagr}) and
therefore can be completely determined in terms of the two free spin-0 fields
given by  Eq.(\ref{field}) and the mixing angle $\theta$.

This, however, gives rise to highly nontrivial
relationship between the Fock-space of free-fields 
and that of interacting fields. 
To build the Fock space of flavor-eigenstates we consider the 
representation of the
transformation consistent with Eq.(\ref{transf}) in the Fock space of mass-eigenstates. 
Using the Baker-Hausdorff lemma,
we can write the generator of this transformation as
\begin{eqnarray}
\hat{S}=&\int& d^3x(\dot{\phi}_{\alpha}^{\dag}(x)\phi_{\beta}(x)+
\phi_{\beta}^{\dag}(x)\dot{\phi}_{\alpha}(x)-
\dot{\phi}_{\beta}^{\dag}(x)\phi_{\alpha}(x)-
\phi_{\alpha}^{\dag}(x)\dot{\phi}_{\beta}(x))\cr
=&\int&
d^3x(\dot{\varphi}_{1}^{\dag}(x)\varphi_{2}(x)+
\varphi_{2}^{\dag}(x)\dot{\varphi}_{1}(x)-
\dot{\varphi}_{2}^{\dag}(x)\varphi_{1}(x)-
\varphi_{1}^{\dag}(x)\dot{\varphi}_{2}(x))
\end{eqnarray}
so that Eq.(\ref{transf}) can be written in the form:
\begin{eqnarray}\label{tr1}
\phi_{\alpha}&=&e^{-i\hat{S}\theta}\varphi_{1}e^{i\hat{S}\theta},\cr
\phi_{\beta}&=&e^{-i\hat{S}\theta}\varphi_{2}e^{i\hat{S}\theta}
\end{eqnarray}
with the transformation operator given by
\begin{equation}\label{gen}
G(\theta)=e^{i\hat{S}\theta}.
\end{equation}
The similarity transformation given by Eq.(\ref{tr1}) relates the free
field operators $\varphi_{1,2}$ to the interacting fields
$\phi_{\alpha,\beta}$. These relationships can be obtained from the
requirement of the inner product conservation 
\begin{equation}
<\alpha|\phi|\beta>=<a|\varphi|b>,
\end{equation}
where a linear transformation  of state vector according to
\begin{equation}
|\alpha>=G^{-1}(\theta)|a>
\end{equation}
relates the two Hilbert spaces:{\it i.e.} mass-eigenstate space ${\cal 
H}_{1,2}$ and flavor-eigenstate space
${\cal H}_{\alpha,\beta}=G^{-1}(\theta){\cal H}_{1,2}$. The 
transformation given by Eq.(\ref{tr1}) can also be viewed as the
"rotation" of basis in the Hilbert space of quantum states
diagonalizing the bilinear lagrangian given by Eq.(\ref{lagr}). 

The operator $\hat{S}$ can then be written in terms of ladder operators
$a_{\vec{k}i}$ and $b_{\vec{k}i}$ as follows:
\begin{eqnarray}\label{gener}
\hat{S}=\sum_{\vec{k}}&\frac{i}{2}
&\big\{\gamma_{+}(a_{\vec{k}1}a_{\vec{k}2}^{\dag}+b_{-\vec{k}1}b_{-\vec{k}2}^{\dag}
-a_{\vec{k}1}^{\dag}a_{\vec{k}2}-b_{-\vec{k}1}^{\dag}b_{-\vec{k}2})+\cr
&+&\gamma_{-}(a_{\vec{k}1}b_{-\vec{k}2}+a_{\vec{k}2}b_{-\vec{k}1}
-a_{\vec{k}1}^{\dag}b_{-\vec{k}2}^{\dag}-b_{-\vec{k}1}^{\dag}a_{\vec{k}2}^{\dag})\big\},
\end{eqnarray}
where we denote $\gamma_{+}=\sqrt{\frac{\epsilon_{1}(k)}
{\epsilon_{2}(k)}}+\sqrt{\frac{\epsilon_{2}(k)}{\epsilon_{1}(k)}}$
and
$\gamma_{-}=\sqrt{\frac{\epsilon_{1}(k)}
{\epsilon_{2}(k)}}-\sqrt{\frac{\epsilon_{2}(k)}{\epsilon_{1}(k)}}$. 
Here $\gamma_{+}^2-\gamma_{-}^2=4$. From Eq.(\ref{gener}) we note that
Eq.(\ref{tr1}) makes each cluster $\Omega_{\vec{k}}$, 
defined by linear superposition of operators
$(a_{\vec{k}1},a_{\vec{k}2},
b_{-\vec{k}1}^{\dag},b_{-\vec{k}2}^{\dag})$, transform into
itself, i.e.
$e^{-i\hat{S}\theta}\Omega_{\vec{k}}e^{i\hat{S}\theta}=\Omega_{\vec{k}}$. 
The same can be said about its hermitian
conjugate $\Omega_{\vec{k}}^{\dag}$. This means that we can consider the
transformation given by Eq.(\ref{tr1}) within each
cluster with a specific momentum $\vec{k}$:
\begin{eqnarray}
\hat{S_{\vec{k}}}=&\frac{i}{2}&\big\{\gamma_{+}
(a_{\vec{k}1}a_{\vec{k}2}^{\dag}+b_{-\vec{k}1}b_{-\vec{k}2}^{\dag}
-a_{\vec{k}1}^{\dag}a_{\vec{k}2}-b_{-\vec{k}1}^{\dag}b_{-\vec{k}2})+\cr
&+&\gamma_{-}(a_{\vec{k}1}b_{-\vec{k}2}+a_{\vec{k}2}b_{-\vec{k}1}
-a_{\vec{k}1}^{\dag}b_{-\vec{k}2}^{\dag}-
b_{-\vec{k}1}^{\dag}a_{\vec{k}2}^{\dag})\big\}.
\end{eqnarray}
Thus, the total transformation is given by 
$e^{i\hat{S}\theta}=\prod_{\vec{k}}e^{i\hat{S}_{\vec{k}}\theta}$. 
It is also convenient to express $\hat{S}_{\vec{k}}$ as
$\hat{S}_{\vec{k}}:=\sqrt{2}(\hat{T}_{\vec{k}}^{\dag}+
\hat{T}_{\vec{k}})$ with the operator $\hat{T}_{\vec{k}}$ defined by
\begin{equation}
\hat{T}_{\vec{k}}=-\frac{i}{2\sqrt{2}}(\gamma_{+}(a_{\vec{k}1}^{\dag}a_{\vec{k}2}
-b_{-\vec{k}1}b_{-\vec{k}2}^{\dag})
+\gamma_{-}(a_{\vec{k}1}^{\dag}b_{-\vec{k}2}^{\dag}-a_{\vec{k}2}b_{-\vec{k}1})),
\end{equation}
where the commutation relation $[\hat{T}_{\vec{k}},\hat{T}_{\vec{k}}^{\dag}]=1$ is
satisfied between $\hat{T}_{k}$ and $\hat{T}^{\dag}_{k}$ just the same 
way as the particle creation and annihilation operators satisfy the  
commutation relations. 
With the operators $\hat{T}_{k}$ and $\hat{T}^{\dag}_{k}$, we can directly apply 
Eq.(\ref{gen}) to the mass-eigenstate vacuum and obtain
\begin{equation}\label{vacpert}
|0;\theta,\vec{k}>_{\alpha,\beta}=\sum_{n=0}^{\infty}\frac{(-i\theta\sqrt{2})^n}{n!}
\sum_{l=0}^{n}C_{n}^{l}(\theta)(\hat{T}_{\vec{k}}^{\dag})^{l}\hat{T}_{\vec{k}}^{n-l}|0>_{1,2},
\end{equation}
where $C_{n}^{l}(\theta)$ are the generalized binomial coefficients that 
can be found after appropriate orderings of $\hat{T}$ and 
$\hat{T}^{\dag}$ are carried out. 
In the expression given by Eq.(\ref{vacpert}), one can treat the operators 
$\hat{T}_{\vec{k}}$ ($\hat{T}_{\vec{k}}^{\dag}$) as the 
annihilation (creation) operator of the vacuum fluctuation. 

For simplicity we now suppress the momentum notation in the ladder 
operators and take the flavor vacuum state in the most general form
\begin{equation}\label{vacser}
|0;\theta>=\sum_{n,l,m,k}C_{nlmk}(\theta)
(a_{1}^{\dag})^{n}(a_{2}^{\dag})^{l}(b_{-1}^{\dag})^{m}(b_{-2}^{\dag})^{k}|0>.
\end{equation}
Applying the flavor annihilation operators to this vacuum, we obtain 
an infinite number of coupled linear equations for the coefficients 
$C_{nlmk}(\theta)$ and solve these equations in Appendix A. As shown in 
Appendix A, we find 
\begin{equation}\label{theta-vacuum}
|0,\theta >= {\cal Z} \sum_{n,l}\frac{1}{n!l!}( Z_{11} 
a_1^{\dagger}b_{-1}^{\dagger } +Z_{12}a_1^{\dagger }b_{-2}^{\dagger })^n
(-Z_{11} a_2^{\dagger }b_{-2}^{\dagger }+Z_{12} a_2^{\dagger 
}b_{-1}^{\dagger })^l|0>, 
\end{equation}
where 
${\cal Z}=<0|0,\theta>$ is the normalization factor to be fixed by
$<\theta;0|0;\theta>=1$ and the coefficients $Z_{11}$ and $Z_{12}$ are 
given by
\begin{eqnarray}\label{Zdefs}
Z_{11}&=&\frac{\gamma_{+}\gamma_{-}\sin^2\theta}{4(\cos^2\theta+
\frac{\gamma_{+}^2}{4}\sin^2\theta)}=
\frac{\gamma_{+}\gamma_{-}\sin^2\theta} 
{4(1+\frac{\gamma_{-}^2}{4}\sin^2\theta)}\cr
Z_{12}&=&\frac{-\gamma_{-}\sin2\theta}{4(\cos^2\theta+
\frac{\gamma_{+}^2}{4}\sin^2\theta)}=\frac{-\gamma_{-}
\sin2\theta}{4(1+\frac{\gamma_{-}^2}{4}
\sin^2\theta)}.
\end{eqnarray}
Here, we note that the coefficients $Z_{11}$ and $Z_{12}$ (see 
Eq.(\ref{Zdefs})) can be written as 
\begin{equation}\label{coeff-redef}
Z_{11}=\chi \cdot x, Z_{12}=\chi \cdot y, 
\end{equation}
where
\begin{eqnarray}
\chi &=&\frac{\gamma _{-}\sin\theta}{2\sqrt{1+\frac{\gamma 
_{-}^2\sin^2\theta}{4}}} \nonumber \\ 
x &=& \frac{\gamma _{+}\sin\theta}{2\sqrt{\cos^2\theta +\frac{\gamma 
_{+}^2\sin ^2\theta}{4}}} \nonumber \\
y &=& \frac{-\cos\theta}{\sqrt{\cos^2\theta+\frac{\gamma _{+}^2\sin 
^2\theta}{4}}}.
\end{eqnarray}
Thus, the flavor vacuum state given by Eq.(\ref{theta-vacuum})
can be rewritten as
\begin{equation}
|0,\theta >={\cal Z} \sum_{n,l}\frac{\chi ^{n+l}}{n!l!}
\left( xa_1^{\dagger}b_{-1}^{\dagger }+ya_1^{\dagger }b_{-2}^{\dagger }\right) ^n
\left(-xa_2^{\dagger }b_{-2}^{\dagger }+ya_2^{\dagger }b_{-1}^{\dagger }\right)^l|0>. 
\end{equation}
This result can be further reduced as 
\begin{equation}
|0,\theta >={\cal Z} \sum_{n,l}\frac{\chi ^{n+l}}{n!l!}
(a_1^{\dagger}c_1^{\dagger })^n(a_2^{\dagger }c_2^{\dagger })^l|0>,
\end{equation}
by defining new ladder operators:
\begin{eqnarray}
c_1&=&xb_{-1}+yb_{-2}, \cr
c_2&=&-xb_{-2}+yb_{-1},
\end{eqnarray}
where it is easy to check that $[c_1,c_1^{\dagger }]=x^2+y^2=1$, 
$[c_2,c_2^{\dagger}]=1$, 
$[c_1,c_2^{\dagger}]=0$. 
Now, it is possible to compute directly the value of ${\cal Z}$, because
\begin{equation}
<0,\theta |0,\theta >={\cal Z} ^2\sum_{n,l}\frac{\chi ^{2(n+l)}}{n!^2l!^2}
n!^2l!^2={\cal Z} ^2\left( \sum_n\chi ^{2n}\right) ^2=\frac{{\cal Z}^2}
{(1-\chi ^2)^2}
\end{equation}
and the flavor vacuum is normalized to be one,{\it i.e.} 
$<0,\theta |0,\theta>=1$. Thus, we find 
\begin{equation}\label{norm-factor}
{\cal Z} =1-\chi
^2=\frac 1{1+\gamma _{-}^2\sin ^2\theta/4}. 
\end{equation}
We see that the flavor vacuum state indeed exists in the Fock space of
mass-eigenstates and 
the normalization factor ${\cal Z}$ is finite but less than one for any 
value of $\gamma_- \sin\theta$ in the exact vacuum treatment. 
The same has been obtained in Ref.\cite{Blasson3} solving a differential
equation of $<0|0,\theta>=<0|G(\theta)|0>$. While such method of  
derivation using a differential equation
has been known for some time\cite{6,8,Blasson3,25}, our 
algebraic method presented in this work is a new development. 

This proves then the unitary inequivalence between the two Fock spaces of 
mass and flavor in the infinite
volume limit following the procedure discussed in Ref.\cite{Blasson3}:
\begin{equation}
\lim_{V\rightarrow\infty} <0|0,\theta>_{\alpha,\beta} =
\lim_{V\rightarrow\infty} \exp(\frac{V}{2\pi^3}\int d^3 k 
\ln {\cal Z}) = 0,
\end{equation}
for any time.
While we agree with Ref.\cite{Blasson3} on the point that only the 
infinite volume limit 
can warrant the unitary inequivalence even in the boson case, we note 
that the perturbative expansion of the exact
vacuum in the boson case is dramatically different from the case of fermion. 
Although the normalization factor ${\cal Z}$ is a finite function
for all values of $\gamma _{-}\sin\theta$, we observe that 
this expression given by Eq.(\ref{norm-factor})
has singularity on the complex plane at 
$\gamma _{-}\sin\theta =2i$.
This is in a remarkable difference from the fermion case where the 
corresponding result ${\cal Z}_{fermion} =1-\gamma_{-}^2 \sin^2 \theta/4$ 
doesn't have any singularity on 
the complex plane. Thus, the flavor vacuum $|0,\theta >$ in terms of
series in $\theta $ shall have a critical point and
this would result in the divergence of the Taylor series
expansion for $<0|0,\theta >$ in powers of $\theta$ 
because such expansion only makes sense for small $\theta$ values.
As we explicitly show in Appendix B, the series given by 
Eq.(\ref{vacpert}) is indeed divergent in the
region $\gamma_{-}\theta>2$.
Such divergence doesn't occur in the fermion case. 
We also present some clarifying remarks in Appendix C regarding on the Green 
function method and the aribitrary mass parametrization discussed in the 
previous literatures\cite{8,Blasson3,ref3}.
\setcounter{equation}{0}
\setcounter{figure}{0}
\renewcommand{\theequation}{\mbox{3.\arabic{equation}}}
\renewcommand{\thefigure}{\mbox{3.\arabic{figure}}}

\section{Ladder Operators and Condensations}
In the previous section, we have built the representation of the mixing 
transformation given by Eq.(\ref{transf}) in the operator space of
$\varphi_{1,2}$, where the action of mixing is given by the similarity 
transformation given by Eq.(\ref{tr1}). We also considered the 
representation defined by operating
$G^{-1}(\theta)$ in the Fock space and showed the unitary
inequivalence between the two (mixed and unmixed) Fock spaces in the 
infinite volume limit.

Let us now further investigate these representations to 
come up with physically measurable quantities.
The fields $\varphi_{1,2}$ are defined by a superposition of ladder operators
$a_{1,2}$ and $b_{1,2}$ that form the basis in a linear Hilbert space of
mass eigenstate fields. Using Eqs.(\ref{field}) and (\ref{tr1}), 
one can immediately obtain annihilation operators for the mixed (flavor)
fields that are consistent with the Pontecorvo mixing relationship;
\begin{eqnarray}\label{annihilation}
a_{\alpha,\beta}&=&G^{-1}(\theta)a_{1,2}G(\theta),\cr
b_{\alpha,\beta}&=&G^{-1}(\theta)b_{1,2}G(\theta).
\end{eqnarray}
This is also consistent with the
definition of flavor vacuum as the lowest energy state,{\it i.e.}
\begin{equation}
<0,\theta|{\hat H}(\theta)|0,\theta>=
<0|G(\theta){\hat H}(\theta)G^{-1}(\theta)|0>=<0|{\hat H}_{0}|0>=0,
\end{equation}
where ${\hat H}(\theta)$ and ${\hat H}_0$ are the Hamiltonians of
mixed-fields and unmixed-fields, respectively.
Straightforward application of Baker-Hausdorff lemma to 
Eq.(\ref{annihilation}) yields: 
\begin{eqnarray}\label{ladden}
a_{\alpha}&=&a_{1}\cos\theta+\frac{\sin\theta}{2}(\gamma_{+}a_{2}+\gamma_{-}b_{-2}^{\dag}),\cr
a_{\beta}&=&a_{2}\cos\theta+\frac{\sin\theta}{2}(-\gamma_{+}a_{1}+\gamma_{-}b_{-1}^{\dag}),\cr
b_{-\alpha}&=&b_{-1}\cos\theta+\frac{\sin\theta}{2}(\gamma_{+}b_{-2}+\gamma_{-}a_{2}^{\dag}),\cr
b_{-\beta}&=&b_{-2}\cos\theta+\frac{\sin\theta}{2}(-\gamma_{+}b_{-1}+\gamma_{-}a_{1}^{\dag}).
\end{eqnarray}
It is also not difficult to reverse Eq.(\ref{ladden}) in order to obtain how the 
mass-eigenstate ladder operators
are expressed in terms of flavor ones. Using the above relationships we can also
find the time dependence of the flavor-eigenstate ladder operators in the 
Heisenberg picture since the time
evolution of mass-eigenstate ladder operators are given by:
\begin{eqnarray}
a_{1,2}(t)&=&e^{i{\hat H}_0 t}a_{1,2}e^{-i{\hat H}_0 t}=
e^{-i\epsilon_{1,2}t}a_{1,2},\cr 
b_{1,2}(t)&=&e^{i{\hat H}_0 t}
b_{1,2}e^{-i{\hat H}_0 t}=e^{-i\epsilon_{1,2}t}b_{1,2}. 
\end{eqnarray}
In particular, after introducing more compact notation
\begin{equation}
C=\cos\theta ;S_{+}=\frac{\sin\theta \gamma_{+}}{2};
S_{-}=\frac{\sin\theta \gamma_{-}}{2},
\end{equation}
we find
\begin{eqnarray}\label{laddentime}
a_{\alpha t}=& &\big(C^2 e^{-i\epsilon_{1}t}+S_{+}^2
e^{-i\epsilon_{2}t}-S_{-}^2e^{i\epsilon_{2}t}\big)a_{\alpha}+
C S_{+}\big(e^{-i\epsilon_{2}t}-e^{-i\epsilon_{1}t}\big)a_{\beta}+\cr
& &S_{+} S_{-}\big(e^{i\epsilon_{2}t}-e^{-i\epsilon_{2}t}\big)
b_{-\alpha}^{\dag}+C S_{-}\big(e^{i\epsilon_{2}t}-
e^{-i\epsilon_{1}t}\big)b_{-\beta}^{\dag},\cr
a_{\beta t}=& &\big(C^2e^{-i\epsilon_{2}t}+S_{+}^2
e^{-i\epsilon_{1}t}-S_{-}^2e^{i\epsilon_{1}t}\big)a_{\beta}+
C S_{+}\big(e^{-i\epsilon_{2}t}-e^{-i\epsilon_{1}t}\big)a_{\alpha}+\cr
& &S_{+}S_{-}\big(e^{-i\epsilon_{1}t}-e^{i\epsilon_{1}t}\big)
b_{-\beta}^{\dag}+C S_{-}\big(e^{i\epsilon_{1}t}-
e^{-i\epsilon_{2}t}\big)b_{-\alpha}^{\dag},\cr
b_{-\alpha t}=& &\big(C^2e^{-i\epsilon_{1}t}+S_{+}^2
e^{-i\epsilon_{2}t}-S_{-}^2e^{i\epsilon_{2}t}\big)b_{-\alpha}+
C S_{+}\big(e^{-i\epsilon_{2}t}-e^{-i\epsilon_{1}t}\big)b_{-\beta}+\cr
& &S_{+}S_{-}\big(e^{i\epsilon_{2}t}-e^{-i\epsilon_{2}t}\big)
a_{\alpha}^{\dag}+C S_{-}\big(e^{i\epsilon_{2}t}-
e^{-i\epsilon_{1}t}\big)a_{\beta}^{\dag},\cr
b_{-\beta t}=& &\big(C^2e^{-i\epsilon_{2}t}+S_{+}^2
e^{-i\epsilon_{1}t}-S_{-}^2e^{i\epsilon_{1}t}\big)b_{-\beta}+
C S_{+}\big(e^{-i\epsilon_{2}t}-e^{-i\epsilon_{1}t}\big)b_{-\alpha}+\cr
& &S_{+}S_{-}\big(e^{-i\epsilon_{1}t}-e^{i\epsilon_{1}t}\big)
a_{\beta}^{\dag}+C S_{-}\big(e^{i\epsilon_{1}t}-
e^{-i\epsilon_{2}t}\big)a_{\alpha}^{\dag},
\end{eqnarray}
from which we can also obtain the \underline{unequal}-time commutation 
relationships:
\begin{eqnarray}\label{commtime}
[a_{\alpha},a_{\alpha t}^{\dag}]&=&[b_{-\alpha},b_{-\alpha t}^{\dag}]
=C^2e^{i\epsilon_{1}t}+S_{+}^2
e^{i\epsilon_{2}t}-S_{-}^2e^{-i\epsilon_{2}t}=A_{\alpha\alpha},\cr
[a_{\beta},a_{\beta t}^{\dag}]&=&[b_{-\beta},b_{-\beta t}^{\dag}]
=C^2e^{i\epsilon_{2}t}+S_{+}^2
e^{i\epsilon_{1}t}-S_{-}^2e^{-i\epsilon_{1}t}=A_{\beta\beta},\cr
[a_{\beta},a_{\alpha t}^{\dag}]&=&[a_{\alpha},a_{\beta t}^{\dag}]=
[b_{-\beta},b_{-\alpha t}^{\dag}]\cr
&=&[b_{-\alpha},b_{-\beta
t}^{\dag}]=C S_{+}\big(e^{i\epsilon_{2}t}-e^{i\epsilon_{1}t}\big)
=A_{\beta\alpha},\cr
[b_{-\beta},a_{\alpha t}]&=&[a_{\beta},b_{-\alpha t}]=-[b_{-\alpha},a_{\beta
t}]^{*}\cr
&=&-[a_{\alpha},b_{-\beta t}]^{*}=
C S_{-}\big(e^{i\epsilon_{2}t}-e^{-i\epsilon_{1}t}\big)=
A_{\bar{\beta}\alpha},\cr
[b_{-\alpha},a_{\alpha t}]&=&[a_{\alpha},b_{-\alpha
t}]=S_{+}S_{-}\big(e^{i\epsilon_{2}t}-e^{-i\epsilon_{2}t}\big)=
A_{\bar{\alpha}\alpha},\cr
[b_{-\beta},a_{\beta t}]&=&[a_{\beta},b_{-\beta t}]=S_{+}S_{-}
\big(e^{-i\epsilon_{1}t}-e^{i\epsilon_{1}t}\big)=A_{\bar{\beta}\beta}.
\end{eqnarray}
All other commutators are either zeros or can be expressed 
in terms of the above ones. 
Eqs.(\ref{ladden}),(\ref{laddentime}),(\ref{commtime}) in fact define
all the dynamics of Pontecorvo mixing for two quantum fields.
To show how these relationships can be used to
calculate the dynamical parameters of the mixed (interacting) fields, one can
consider the time evolution of cluster $\Omega_{\vec{k}}$ defined in 
Section II. As discussed in Section II, however, this cluster is
invariant under the $G^{-1}(\theta)$ transformation. Thus, we can consider
$\Omega_{\vec{k}}$ with a particular $\vec{k}$ independently from all other 
momentum values.

We now calculate the number of particles with a definite mass condensed in the
flavor vacuum state $|0'>=|0>_{\alpha,\beta}$. Let's consider the 
condensation of the particle with a definite mass, for example
$Z_{1}=<0'|N_{1}|0'>$. Using the inverse relation of Eq.(\ref{ladden}):
\begin{equation}
a_{1}=a_{\alpha}\cos\theta -\frac{\sin\theta}{2}
(\gamma_{+}a_{\beta}+\gamma_{-}b_{-\beta}^{\dag}),
\end{equation}
we can get
\begin{equation}\label{z1z2}
Z_1=<0'|a_{1}^{\dag}a_{1}|0'>=\frac{\sin^2\theta\gamma_{-}^2}
{4}<0'|b_{-\beta}b_{-\beta}^{\dag}|0'>.
\end{equation}
One can show that the same result is true for $Z_{2}=<0'|N_{2}|0'>$. 
Thus, the condensate density of particles with a definite mass 
in the flavor vacuum is given by:
\begin{equation}\label{condensate}
Z_1=Z_2=S_{-}^2=\frac{\sin^2\theta\gamma_{-}^2}{4}.
\end{equation}
Apparently the condensate densities for particles with definite flavor 
in the mass vacuum, 
{\it i.e.} $<0|N_{\alpha(\beta)}|0>$, are also given by $S_{-}^2$.
Let us now consider the number of 
particles with a definite flavor in the flavor vacuum,
for example $Z_{\alpha}(t)=<0'|N_{\alpha}(t)|0'>$. 
Using Eq.(\ref{laddentime}), one can easily show that
\begin{eqnarray}
Z_{\alpha}(t)=<0'|\big(S_{+}S_{-}\big(e^{i\epsilon_{2}t}-e^{-i\epsilon_{2}t}\big)
b_{-\alpha}^{\dag}+C S_{-}\big(e^{i\epsilon_{2}t}-
e^{-i\epsilon_{1}t}\big)b_{-\beta}^{\dag}\bigg)^{\dag}\cdot\cr
\big(S_{+} S_{-}\big(e^{i\epsilon_{2}t}-e^{-i\epsilon_{2}t}\big)
b_{-\alpha}^{\dag}+C S_{-}\big(e^{i\epsilon_{2}t}-
e^{-i\epsilon_{1}t}\big)b_{-\beta}^{\dag}\big)|0'>
\end{eqnarray}
and thus 
\begin{equation}\label{flavcondensate}
Z_{\alpha}(t)=4S_{-}^2S_{+}^2\sin^2(\epsilon_{2}t)+
4S_{-}^2C^2\sin^2(\frac{\epsilon_{1}+\epsilon_{2}}{2}t).
\end{equation}
Similarly, we get for the $\beta$-particles
\begin{equation}\label{flavcondensate1}
Z_{\beta}(t)=4S_{-}^2S_{+}^2\sin^2(\epsilon_{1}t)+
4S_{-}^2C^2\sin^2(\frac{\epsilon_{1}+\epsilon_{2}}{2}t).
\end{equation}
We see that the number of particles with a definite flavor 
in the flavor vacuum is indeed not 
zero. This is due to the fact that flavor vacuum is not an energy 
eigenstate of the hamiltonian ${\hat H}(\theta)$
and changes with the time translation producing and destroying coherently 
virtual particle/antiparticle pairs. It shows a significant difference 
from the ordinary quantum 
mechanical treatment without considerating the vacuum effect,
which yields $Z_{\alpha (\beta)}=0$ for any time.
We emphasize that our flavor vacuum here is not 
perturbative but exact. This is different
from the approach, where mass eigenstate vacuum 
$|0>_{1,2}$ is used instead of the flavor vacuum
to generate a flavor eigenstate,{\it e.g.} 
$|\alpha>=a_{\alpha}^{\dag}|0>_{1,2}$. 
If the flavor vacuum $|0^\prime>$ was replaced by the mass vacuum 
$|0>_{1,2}$, then we would have obtained $Z_1 = Z_2 = 0$ instead of
Eq.(\ref{condensate}). As discussed above in the exact vacuum treatment,  
the mass eigenstate vacuum is not annihilated by $a_{\alpha,\beta}$ 
operators. Indeed the term proportional to $O(\gamma_{-})$ remains in the 
creation/annihilation operators, so that the accuracy in 
the order of $O(\gamma_{-}^2)$ can be expected from the results of  
exact vacuum approach compare to the perturbative vacuum approximation.
The densities of vacuum condensation for antiparticles
$Z_{\bar{1},\bar{2}},Z_{\bar{\alpha},\bar{\beta}}$ are obtained same as the 
densities for the corresponding particles, i.e. 
$Z_{1,2},Z_{\alpha,\beta}$ given by
Eqs.(\ref{condensate}),(\ref{flavcondensate}) and (\ref{flavcondensate1}).

We now consider the flavor oscillations in time
for a single particle with flavor $\alpha$ and momentum $\vec{k}$.
In the Heisenberg picture, the average
number of particles with flavor $a = \alpha$ or $\beta$ in the 
flavor state $|\alpha> = a^\dag_\alpha |0^\prime>$
is given by:
\begin{equation}\label{eqq1}
<N_{a}(t)>=<\alpha|N_{a}(t)|\alpha>=<0'|a_{\alpha}a_{a t}^{\dag}a_{a
t}a_{\alpha}^{\dag}|0'>.
\end{equation}
In Eq.(\ref{eqq1}), we note that we use the flavor vacuum to obtain 
exact result for the flavor oscillations. Later, in Section IV, we 
numerically compare the exact result with the previous approximate 
result\cite{7}. 
Using Eqs.(\ref{ladden}),(\ref{laddentime}),(\ref{commtime}),
we directly apply the standard quantum field theoretic method.
Since the flavor vacuum is annihilated by $a_{\alpha,\beta}$, 
we move $a_{\alpha}$ in Eq.(\ref{eqq1})
to the most right position and $a_{\alpha}^{\dag}$ to the most left position
to annihilate the flavor vacuum. 
What is left is uniquely determined by the unequal time commutation
relations given by Eq.(\ref{commtime}) and we find
\begin{eqnarray}\label{313}
<\alpha|N_{\alpha t}|\alpha>&=&<0'|a_{\alpha t}^{\dag}a_{\alpha
t}|0'>+|[a_{\alpha},a_{\alpha t}^{\dag}]|^2=
Z_{\alpha}+|A_{\alpha\alpha}|^2;\cr
<\alpha|N_{-\bar{\alpha} t}|\alpha>&=&<0'|b_{-\alpha t}^{\dag}b_{-\alpha t}|0'>+
|[a_{\alpha},b_{-\alpha t}]|^2=
Z_{\alpha}+|A_{\bar{\alpha}\alpha}|^2;\cr
<\alpha|N_{\beta t}|\alpha>&=&<0'|a_{\beta t}^{\dag}a_{\beta
t}|0'>+|[a_{\alpha},a_{\beta t}^{\dag}]|^2=
Z_{\beta}+|A_{\beta\alpha}|^2;\cr
<\alpha|N_{-\bar{\beta} t}|\alpha>&=&<0'|b_{-\beta t}^{\dag}b_{-\beta 
t}|0'>+ |[a_{\alpha},b_{-\beta t}]|^2=
Z_{\beta}+|A_{\bar{\beta}\alpha}|^2.
\end{eqnarray}
Using the notation of $C$, $S_{\pm}$, our results are summarized as:
\begin{eqnarray}\label{N_time}
<\alpha|N_{\alpha
t}|\alpha>&=&1+8 C^2 S_{-}^2\sin^2(\frac{\epsilon_{1}+\epsilon_{2}}{2}t)
+8 S_{-}^2 S_{+}^2\sin^2(\epsilon_{2}t)-4 C^2 S_{+}^2\sin^2(\frac{\epsilon_{1}-\epsilon_{2}}{2}t),\cr
<\alpha|N_{\beta
t}|\alpha>&=&4 C^2 S_{-}^2\sin^2(\frac{\epsilon_{1}+\epsilon_{2}}{2}t)
+4 S_{-}^2 S_{+}^2\sin^2(\epsilon_{1}t)+4 C^2 S_{+}^2\sin^2(\frac{\epsilon_{1}-\epsilon_{2}}{2}t),\cr
<\alpha|N_{-\bar{\alpha}
t}|\alpha>&=&4 C^2 S_{-}^2\sin^2(\frac{\epsilon_{1}+\epsilon_{2}}{2}t)
     +8 S_{-}^2 S_{+}^2\sin^2(\epsilon_{2}t),\cr
<\alpha|N_{-\bar{\beta}
t}|\alpha>&=&8 C^2 S_{-}^2\sin^2(\frac{\epsilon_{1}+\epsilon_{2}}{2}t)
     +4 S_{-}^2 S_{+}^2\sin^2(\epsilon_{1}t).
\end{eqnarray}
As shown in Eq.(\ref{N_time}),the time dependence of the average number 
of particles with a definite flavor is rather complicate.
It contains oscillating contributions both from
the $\alpha\rightarrow\beta$ conversion and from the virtual pair 
creation in a dynamically "rotating" flavor vacuum. 
As discussed in Ref.\cite{7}, the $\alpha\rightarrow\beta$ conversion
process generates the term proportional to $\sin^2(\frac{\epsilon_1 - 
\epsilon_2}{2} t)$. 
The terms involving  
$\epsilon_{1}+\epsilon_{2},\epsilon_{1},\epsilon_{2}$-frequencies 
in Eq.(\ref{N_time}) are, however, related to the creation of virtual pairs. 
For example, the virtual pair 
creation violates energy conservation
within the uncertainty time,{\it i.e.} $\Delta E\Delta t\approx 1$ (in our 
units $\hbar=1$) and thus both creation and annihilation of, let's say, 
($\alpha+\bar{\beta}$) virtual pair must occur within
$\tau\approx\frac{1}{\epsilon_{1}+\epsilon_{2}}$ time interval.
Thus, the terms in Eq.(\ref{N_time}) involving
$\epsilon_{1}+\epsilon_{2},\epsilon_{1},\epsilon_{2}$-frequencies can be
related to the creation of different types of
virtual pairs, while the terms involving $\epsilon_{1}-\epsilon_{2}$ is
related to the actual $\alpha\rightarrow\beta$ conversion. 

Using Eq.(\ref{N_time}),
we can also calculate the
expectation value of the flavor charge operator defined by 
$Q_{\alpha,\beta}= N_{\alpha,\beta}-N_{-\bar{\alpha},-\bar{\beta}}$;
\begin{eqnarray}\label{Q_time}
<Q_{\alpha}>&=&1-4 C^2 S_{+}^2\sin^2(\frac{\epsilon_{1}-\epsilon_{2}}{2}t)
+4 C S_{-}^2 \sin^2(\frac
{\epsilon_{1}+\epsilon_{2}}{2}t),\cr
<Q_{\beta}>&=&4 C^2 S_{+}^2\sin^2(\frac{\epsilon_{1}-\epsilon_{2}}{2}t)-4 C^2 S_{-}^2\sin^2(\frac
{\epsilon_{1}+\epsilon_{2}}{2}t),
\end{eqnarray}
or with the conventional parameters,
\begin{eqnarray}\label{resfin}
<Q_{\alpha}>&=&1-\gamma_{+}^2\sin^2(2\theta)\sin^2(\frac{\epsilon_{1}-\epsilon_{2}}{2}t)+
\gamma_{-}^2\sin^2(2\theta)\sin^2(\frac{\epsilon_{1}+\epsilon_{2}}{2}t),\cr
<Q_{\beta}>&=&\gamma_{+}^2\sin^2(2\theta)\sin^2(\frac{\epsilon_{1}-\epsilon_{2}}{2}t)-
\gamma_{-}^2\sin^2(2\theta)\sin^2(\frac{\epsilon_{1}+\epsilon_{2}}{2}t).
\end{eqnarray}
From this result, one can also see that there is an additional term
proportional to $\sin^2(\frac{\epsilon_{1}+\epsilon_{2}}{2}t)$ 
to the usual Pontecorvo formula. As discussed above, the origin of
this term can be understood as a contribution from the virtual pair 
creation in "rotating" vacuum. The correction term is of the order of 
$O(\gamma_{-}^2)$. As noted earlier, this may give a reason why it 
has been found neither in an ordinary quantum mechanical treatment nor 
in the approximate QFT treatment based on a perturbative vacuum.

We also calculate the time evolution of coherent state for the two mixed 
quantum fields.
Coherent state has the form
\begin{equation}
|C\alpha>=e^{C a_{\alpha}^{\dag}}|0'>.
\end{equation}
Extending the above calculation for a single particle, it is not so 
difficult to verify that for the state containing $n$ particles with
flavor $\alpha$ can be given by
\begin{equation}\label{n-alpha}
<n|N_{\alpha t}|n>=\frac{1}{n!}<0^\prime|a_{\alpha}^n N_{\alpha 
t}(a_{\alpha}^{\dag})^n|0^\prime>= <N_{\alpha t}>+n |A_{\alpha\alpha}|^2.
\end{equation}
Besides $n |A_{\alpha \alpha}|^2$ which is simply $n$ times 
the probability of $\alpha\rightarrow\alpha$ transition, we see in 
Eq.(\ref{n-alpha}) that the condensate contribution is present adding the
density of $\alpha$ particles from "rotating" vacuum.
Applying this result directly to the coherent state expansion,  we obtain 
the following expectation values of the number operator 
$N_{(\alpha,\beta)t}$ in the coherent state $|C\alpha>$:
\begin{eqnarray}
<C\alpha|N_{\alpha t}|C\alpha>=Z_{\alpha}+|C|^2|A_{\alpha\alpha}|^2,\cr
<C\alpha|N_{\beta t}|C\alpha>=Z_{\beta}+|C|^2|A_{\beta\alpha}|^2.
\end{eqnarray}
Thus, the expectation values of the flavor charge operator 
$Q_{(\alpha,\beta)} = N_{(\alpha,\beta)}-N_{(-{\bar \alpha},-{\bar \beta})}$
turn out to be
\begin{eqnarray}\label{coherent-charge}
<C\alpha|Q_{\alpha}|C\alpha>&=&|C|^2<Q_{\alpha}>\cr
          &=&|C|^2(1-\gamma_{+}^2\sin^2(2\theta)
             \sin^2(\frac{\epsilon_{1}-\epsilon_{2}}{2}t)+
\gamma_{-}^2\sin^2(2\theta)\sin^2(\frac{\epsilon_{1}+\epsilon_{2}}{2}t)),
\nonumber \\
<C\alpha|Q_{\beta}|C\alpha>&=&|C|^2<Q_{\beta}>\cr
            &=&|C|^2(\gamma_{+}^2\sin^2(2\theta)
               \sin^2(\frac{\epsilon_{1}-\epsilon_{2}}{2}t)-
\gamma_{-}^2\sin^2(2\theta)\sin^2(\frac{\epsilon_{1}+\epsilon_{2}}{2}t)).
\end{eqnarray}
As we can see in Eq.(\ref{coherent-charge}),
the vacuum contributions $Z_{(\alpha,\beta)}$ are removed from 
the flavor charge expectation values and the results for the coherent 
state are simply $|C|^2$ times the expectation values of the flavor 
charge for the single particle state.

\setcounter{equation}{0}
\setcounter{figure}{0}
\renewcommand{\theequation}{\mbox{5.\arabic{equation}}}
\renewcommand{\thefigure}{\mbox{5.\arabic{figure}}}

\section{Application to Real Meson States }
We now apply the results for time evolution of two mixing boson
fields to the analysis of $\eta-\eta'$ mixing system.
The masses are taken to be 549 MeV and 958 MeV, respectively, and
of course in the particle rest frame the energies in our formulas
reduce to the masses. The phenomenologically allowed mixing angle
($\theta_{SU(3)}$) range of the $\eta {\eta}'$ system is given between
$-10^\circ$ and $-23^\circ$\cite{PDG}, where the mixing angle
$\theta_{SU(3)}$ is defined by Eq.(36) of Ref.\cite{CJ}. This angle
represents the breaking of the SU(3) symmetry, the eigenstates of
which are already rotated $-35.26^\circ$ from $u {\bar u} +d {\bar d}$
and $s {\bar s}$ to $\alpha = u {\bar u} + d {\bar d} -2 s {\bar s}$
and $\beta = u {\bar u} + d {\bar d} + s {\bar s}$. Thus, our mixing
angle is defined by $\theta = \theta_{SU(3)} - 35.26^\circ$.
Recent analysis of the $\eta {\eta}'$ mixing angle using a constituent
quark model based on the Fock states quantized on the light-front can be
found in Ref.\cite{CJ} and the references therein. The optimal value
found for $\theta_{SU(3)}$ was around $-19^\circ$ and thus
$\theta \approx -54^{\circ}$.
We use these values in Eqs.(\ref{N_time}) and (\ref{Q_time}) (or 
equivalently 
(\ref{resfin})) to determine the evolution of definite flavor particle 
number and charge.

\begin{figure}
\begin{center}
\epsfig{bbllx=100, bblly=100, bburx=550, bbury=550,
angle=-90,width=250pt, file=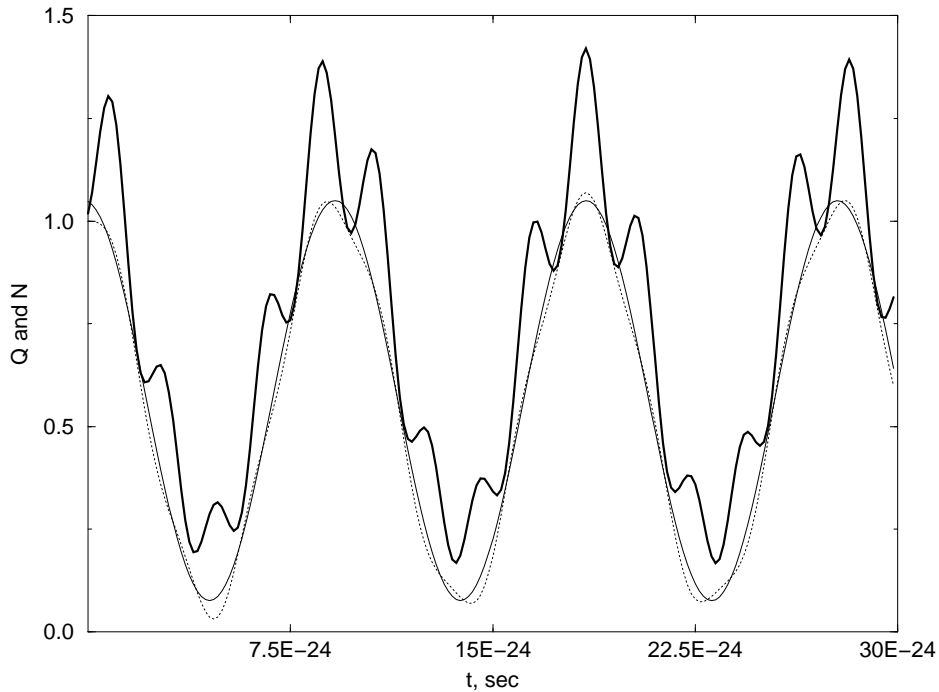}
\end{center}
\caption{comparing population density evolution for k=0.1GeV}\label{figC1}
\end{figure}

In Fig.\ref{figC1}, we present both $<\alpha| N_{\alpha t} |\alpha>$ (thick 
solid line)
and $<Q_\alpha>$ (dotted line) as a function of time when the particle 
momentum is 
given by $k=0.1$ GeV. For a comparison, we also show the previous approximate
result (thin solid line) based on the perturbative vacuum\cite{7} 
corresponding to these quantities noting that $<\alpha| N_{\alpha t} 
|\alpha>$ and $<Q_\alpha>$ coincide each other in this approximation
as one can see in Eqs.(\ref{N_time}) and (\ref{Q_time}).
As we shown in Fig.\ref{figC1}, the population density 
$<\alpha| N_{\alpha t} |\alpha>$ (thick solid line) is completely
distorted due to the interaction with the nonperturbative vacuum while
the sinusoidal Pontecorvo result (thin solid line) is obtained 
for the approximate perturbative vacuum treatment. We see the large
deviation up to $40\%$ in $<\alpha| N_{\alpha t} |\alpha>$.
However, one cannot see the same level of deviation in
$<Q_\alpha>$ and the previous result\cite{7} based on the perturbative
vacuum seems to be a good approximation for the description of
flavor charge oscillations modulo the accuracy of order 
$O(\gamma_{-}^2)$.

\begin{figure}
\begin{center}
\epsfig{bbllx=100, bblly=100, bburx=550, bbury=550,
angle=-90,width=200pt, file=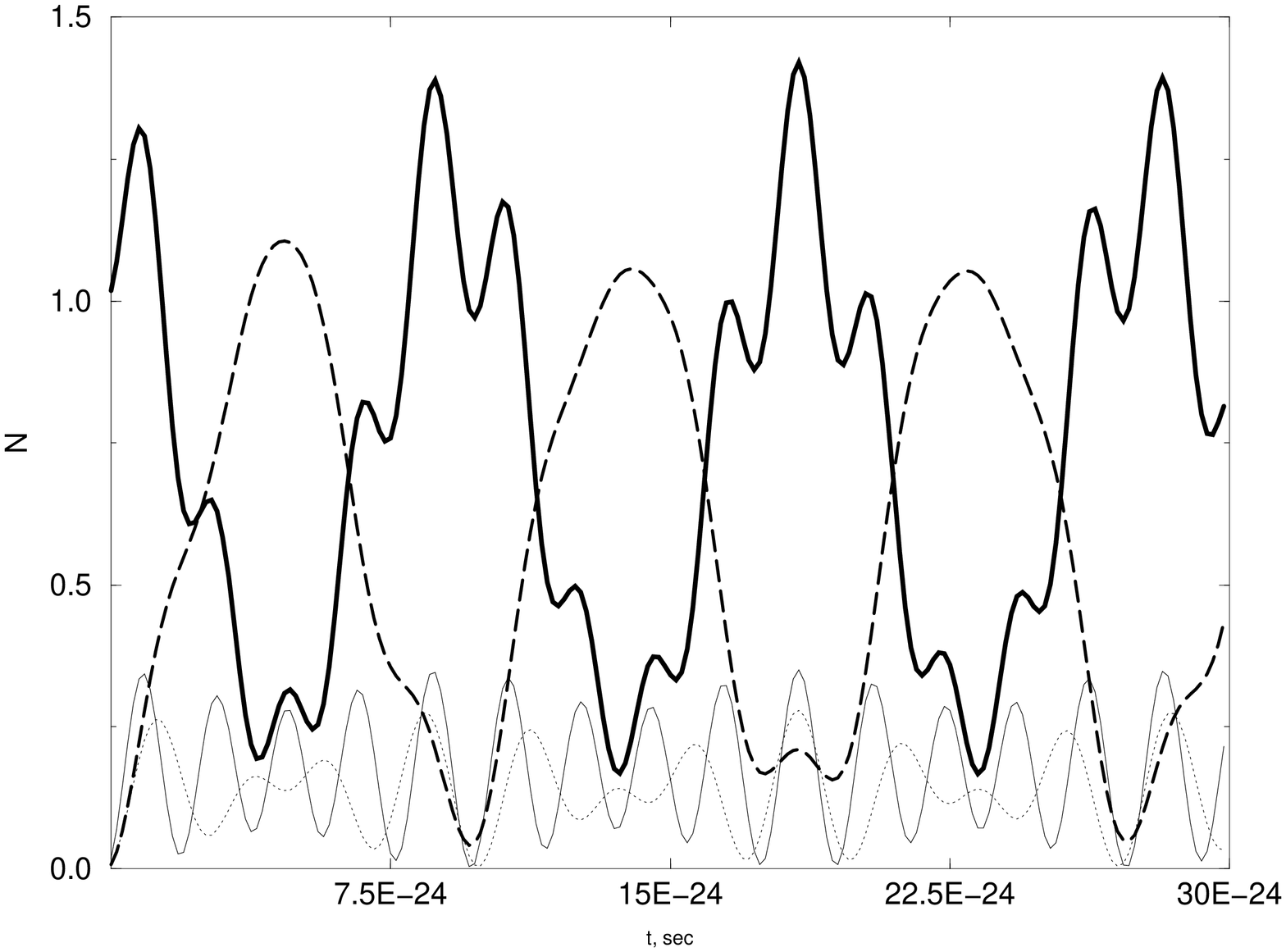}
\end{center}
\caption{population density evolution for k=0.1GeV}\label{figN2}
\end{figure}

More details of our results on the time evolution of the particle 
number with the momentum $k=0.1$ GeV are shown in Fig.\ref{figN2}, where
the thick solid and dashed lines are $<\alpha| N_{\alpha t} |\alpha>$
and $<\alpha| N_{\beta t} |\alpha>$, respectively, and the thin solid
and dotted lines are respectively the antiparticle contributions of
$<\alpha| N_{{\bar \alpha} t} |\alpha>$ and $<\alpha| N_{{\bar \beta} t} 
|\alpha>$.  
The $\eta \eta'$ is one of the most severely mixed systems
due to the great difference in masses of mixed particles.
As we have stated earlier, the simple harmonic structure of average
particle number usually obtained in quantum mechanics or in an 
approximate QFT treatment\cite{7} is completely altered as a result of
nontrivial interaction with the complicate vacuum.
What we see is the superposition of two different cycles as described 
by Eq.(\ref{N_time}). From the initial moment of
time the population of both $\alpha$-particles (thick solid line) 
and $\beta$-particles (thick dashed line) increases. 
Although the increase of number of $\beta$-particles
in system is well understood due to $\alpha\rightarrow\beta$ conversion, 
the initial increase of $\alpha$-population is quite
unexpected and caused by $\alpha-\bar{\alpha}$ production from vacuum. 
The contribution from this process however is rather fast so that the 
general tendency of exchanging between $\alpha$ and
$\beta$ particle states can also be seen quite well. In Fig.\ref{figN2}, 
we also see the oscillations of the antiparticle number in the 
system. This effect is given in the order of $\gamma_{-}^2$ and usually
is absent in an approximate QFT treatment. This is entirely a QFT effect 
which cannot be obtained within the framework of quantum mechanics.  
In QFT, besides the beams of $\alpha$ and $\beta$ particles moving in 
$\vec{k}$ direction, we necessarily have antiparticle beam traveling in 
the opposite direction. The population density in this beam is
correlated with particle-beam so that the total flavor is preserved. 
The existence of beam is caused by "dynamically rotating" vacuum
disturbance at the initial time of $\alpha$-particle emerging.
One should also note that the existence of "recoil" antiparticle beam 
is preserved in the more general wave-packet QFT treatment of
mixing problem. Thus, the mixed particle of definite flavor 
not only produces the usual oscillation of population density in time (or 
space\cite{Lipkin}) but also is accompanied by emitting
the beam of antiparticles traveling in the direction opposite to the 
beam of particles. These effects are in principle testable in the
experiments.

\begin{figure}
\begin{center}
\epsfig{bbllx=100, bblly=100, bburx=550, bbury=550,
angle=-90,width=200pt, file=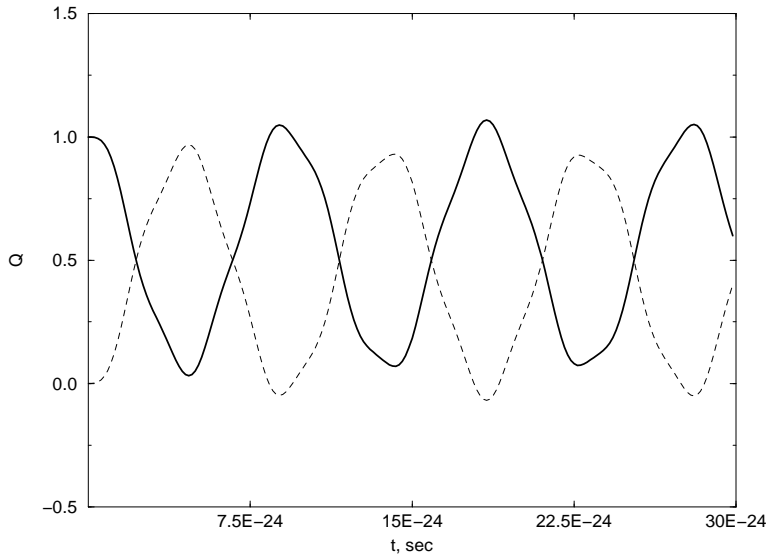}
\end{center}
\caption{flavor charge oscillations}\label{figQ1}
\end{figure}

In Fig.\ref{figQ1}, we also plot more details on the 
time dependence of flavor charge expectation value
with the same momentum $k=0.1$ GeV. The thick solid and dashed
lines are $<Q_\alpha>$ and $<Q_\beta>$, respectively.
One can see that they exhibit mainly the simple periodic structure
similar to the approximate QFT-results
\cite{6,7} and only slightly distorted due to interaction with vacuum. 
The amount of distortion is of $\gamma_{-}^2$ order, {\it i.e.} about 10\% for 
this case. Interesting feature is however
presence of the regions where flavor charge of given sort of particles change
sign which means that
antiparticles outnumber the particles. The process can be physically understood
as result of $\alpha-\bar{\alpha}$
production when number of $\alpha$-particles is small due to
$\alpha\rightarrow\beta$ transition.

\begin{figure}
\begin{center}
\epsfig{bbllx=100, bblly=100, bburx=550, bbury=550,
angle=-90,width=200pt, file=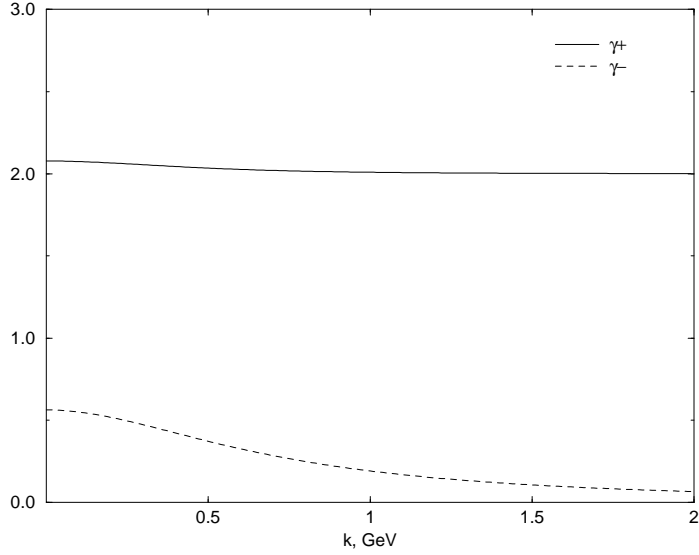}
\end{center}
\caption{mixing amplitudes}\label{figG1}
\end{figure}

It is also interesting and experimentally testable that the efficiency of
conversion processes and the flavor-vacuum disturbance
depend essentially on the energy of original particle. The dependence is
effective to the relativistic  mass of particles so that the
QFT-mixing effects are decreasing with the energy increase of
flavor-particle. The distribution of intensity for simple quantum 
mechanical mixing and QFT-mixing is given by the relationship of amplitudes 
$\gamma_{+}(k)$, $\gamma_{-}(k)$ which determine the intensity of
$a_{2}$ and $b_{-2}^{\dag}$ terms in $a_{\alpha}$ (See 
Eq.(\ref{ladden})). In Fig.\ref{figG1}, we plot their dependence on 
momentum of emitted $\alpha$ particle. As we can see in Fig.\ref{figG1},
$\gamma_{+}$ amplitude falls down as $k$ increases and goes to 2 as 
$k\rightarrow\infty$. In this limit, $\gamma_{+}$  
defines mixing due to a simple rotation between $a_{1}$ and
$a_{2}$ states. Since it can be successfully computed within the 
framework of quantum mechanics, it gives the usual Pontecorvo formula 
with only one oscillatory term. On the other hand, $\gamma_{-}$
appears with an antiparticle creation operator and describe Bogoliubov rotation
between $a_{1}$ and $b_{-1}$ states. This term is also responsible for 
$\frac{\epsilon_{1}+\epsilon_{2}}{2}$ high frequency
term and antiparticle beam creation. As we see in Fig.\ref{figG1}, it
decreases as $k\rightarrow\infty$ and the mass
difference becomes washed out by the relativistic gain of mass. This also means that
at ultrarelativistic limit the 
QFT-mixing effects vanish so that the simple Pontecorvo formula is restored for
flavor-oscillation.

\begin{figure}
\begin{center}
\epsfig{bbllx=100, bblly=100, bburx=550, bbury=550,
angle=-90,width=200pt, file=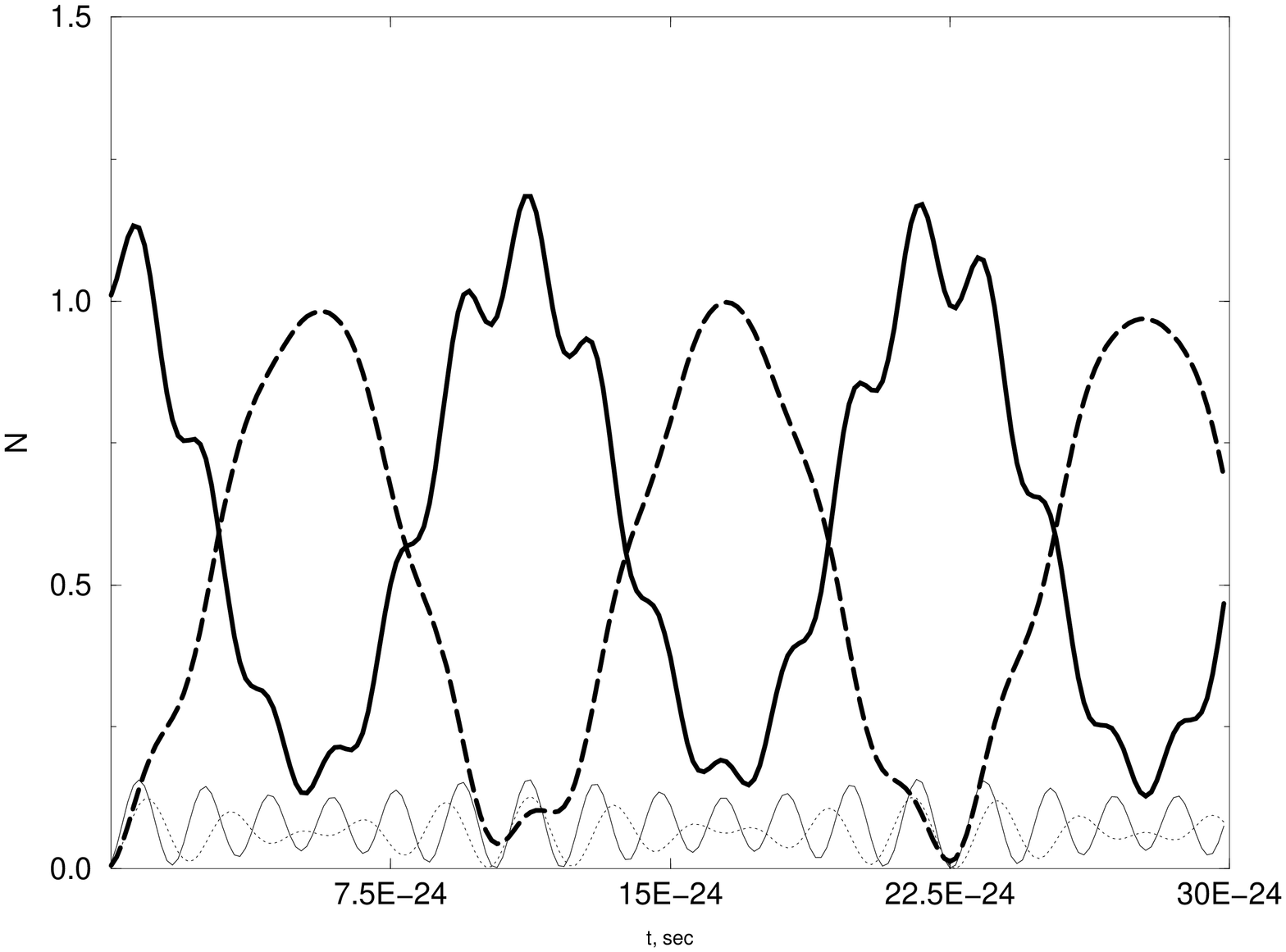}
\end{center}
\caption{population density evolution for k=0.5GeV}\label{figN1}
\end{figure}

To demonstrate the energy dependence, we show in Fig.\ref{figN1}
the plot of population densities evolving with time for the larger
momentum $k=0.5$ GeV.
The line assignments are same as shown in Fig.\ref{figN2}.
As easily seen in Fig.\ref{figN1}, the intensity of antiparticle beam 
decreases dramatically to about 10\%
(in contrast to 20-40\% in Fig.\ref{figN2}) of initial intensity. The 
initial increase in the population density fluctuation in particle beams
also reduces even though the quantum mechanical simple oscillations with
$\frac{\epsilon_{1}-\epsilon_{2}}{2}$
frequency are still visibly distorted. Two beams nevertheless demonstrate 
strong correlation of the same kind as
correlation in quantum mechanical EPR-paradox problem 
so that total flavor charge is
preserved as it should be. It is also noticeable that there exist moments 
of time when the antiparticle outnumbers the particle of the same sort thus 
producing a negative flavor charge as shown in Fig.\ref{figQ1}.

\section{Conclusions and Discussions }
The non-trivial scalar and pseudoscalar meson mixing effects may be
understood by the condensation of corresponding flavor
states in the vacuum{\cite{7}}. We have now extended the analysis
replacing the perturbative vacuum to the nonperturbative (flavor) vacuum.
Central to this analysis is the interplay between the base (unmixed)
Fock space and the physical Fock space. Their nontrivial relationship
%(unitary inequivalence of the vacuum states)
gives rise to the mixing and oscillation phenomena. While the similar 
quantum field theoretic formulation was presented for the fermion mixing 
{\cite{6,8}}, as well as boson mixing\cite{Blasson3}, our analysis
differs in the derivation of the normalization factor ${\cal Z}$ given by
Eq.(\ref{norm-factor}) which is crucial to show the unitary inequivalence
between the mass vacuum and the flavor vacuum.
We presented a new algebraic method which is distinct from the 
conventional method of using a differential equation for ${\cal Z}$.
While the unitary inequivalence occurs only in the infinite volume limit
even for the boson case as discussed in Ref.\cite{Blasson3}, we find 
an intrinsic difference between the fermion and boson cases. As shown in 
this work, the normalization factor ${\cal Z}$ for the boson given by
Eq.(\ref{norm-factor}) has a singularity on the complex plane at 
$\gamma_{-} \sin\theta = 2 i$ while 
the corresponding result for the fermion
%, {\it i.e.} ${\cal Z}_{fermion}
%= 1- \gamma_{-}^2\sin^2\theta/4$, 
doesn't have any singularity.
As we summarized in Appendix B, this singularity corresponds
to the divergence of the Taylor series expansion in powers of $\theta$
for the region $\gamma_{-}\theta > 2$.
For both the boson and fermion cases, however, the non-trivial observable 
mixing phenomena cannot occur unless there is both a nonzero mixing angle and also a
nonzero mass (energy) difference between the two physically
measurable mixed states. Dramatic small oscillations due to the virtual pair 
creation occur in the exact vacuum analysis, while only a simple 
sinusoidal Pontecorvo oscillations occur in the
perturbative vacuum treatment. 
Some clarifying remarks on the Green function method and the arbitrary
mass parametrization discussed in the previous 
literatures\cite{8,Blasson3,ref3} are also summarized in Appendix C.

As a physical application, we used our formulation to analyze the
${\eta}$ ${\eta}'$ system and found that the measured mixing angle and
mass difference between ${\eta}$ and ${\eta}'$ can be related to the
non-trivial flavor condensation in the vacuum.
However, more fundamental questions such as the translation of the
condensation in hadronic degrees of freedom to those in quark and gluon
degrees of freedom remains unanswered. The answer to this question
depends on the dynamics responsible for the confinement of quark and
gluon degrees of freedom and perhaps has to rely on lattice QCD and/or
some phenomenological model that accommodates strongly interacting QCD.
Further investigation along this line is underway. Also, it would be
interesting to look at the mixing transformations between gauge vector
bosons governed by the Weinberg angle in the electroweak theory as well
as vector mesons such as the $\rho$ and $\omega$.
While the statistics are the same as the scalar and pseudoscalar bosons
considered here, there will be additional spin dependent interactions
which complicate the analysis.

{\bf Acknowledgement}\\
\hspace*{\parindent}
This work was supported by the Department of Energy under the contract
DE-FG02-96ER40947. The North Carolina Supercomputing
Center is also acknowledged for the grant of Cray time.
\noindent

\newpage

\setcounter{equation}{0}
\setcounter{figure}{0}
\renewcommand{\theequation}{\mbox{A.\arabic{equation}}}
\renewcommand{\thefigure}{\mbox{A.\arabic{figure}}}

\begin{appendix}
\section{Coupled equations for computing the flavor vacuum 
structure}\label{app1}

In this Appendix, we summarize the procedure for deriving 
Eq.(\ref{theta-vacuum}) that describes a structure of the flavor vacuum.  
We first define flavor vacuum in terms of mass-eigenstates as the most 
general linear superposition of the form:
\begin{eqnarray}
|0;\theta>=\sum_{n,l,m,k}C_{nlmk}(\theta)
(a_{1}^{\dag})^{n}(a_{2}^{\dag})^{l}(b_{-1}^{\dag})^{m}(b_{-2}^{\dag})^{k}|0>
\cr =
\sum_{n,l,m,k}\frac{{C'}_{nlmk}(\theta)}{n!l!}
(a_{1}^{\dag})^{n}(a_{2}^{\dag})^{l}(b_{-1}^{\dag})^{m}(b_{-2}^{\dag})^{k}|0>.
\end{eqnarray}
Then, using the definition of flavor vacuum:
\begin{eqnarray}
&a&_{\alpha,\beta}|0;\theta>=0; \cr
&b&_{-\alpha,-\beta}|0;\theta>=0,
\end{eqnarray}
and explicit expression for the ladder operators given by Eq.(\ref{ladden}),
one can derive an infinite set of linear equations for ${C'}_{nlkm}$ 
coefficients:
\begin{eqnarray}
C {C'}_{n+1,lmk}+S_{+} {C'}_{n,l+1,mk}+S_{-} {C'}_{nlm,k-1}&=&0, \cr
C {C'}_{n,l+1,mk}-S_{+} {C'}_{n+1,lmk}+S_{-} {C'}_{nl,m-1,k}&=&0, \cr
n,l,m,k=0, 1, 2, 3,\dots\cr
\end{eqnarray}
To solve this infinite set of equations, we can express 
${C'}_{n+1,lmk},{C'}_{n,l+1,km}$ in terms of 
${C'}_{nl,m-1,k},{C'}_{nlm,k-1}$ so that we step by step reduce the $n+l$ 
number of particles. Denoting 
\begin{equation}\label{Zdef}
\left(
\begin{array}{cc}
Z_{12} & Z_{11} \\
Z_{22} & Z_{21} 
\end{array}\right)=-S_{-} \left(
\begin{array}{cc}
C & S_{+} \\
-S_{+} & C
\end{array} \right)^{-1}
\end{equation}
(notation for $\hat{Z}$ is chosen in correspondence to 
the index of particle type), we can write this relationship as
\begin{eqnarray}\label{CviaZ}
{C'}_{n+1,lmk}&=&Z_{12}{C'}_{nlm,k-1}+Z_{11}{C'}_{nl,m-1,k} \cr
{C'}_{n,l+1,mk}&=&Z_{22}{C'}_{nlm,k-1}+Z_{21}{C'}_{nl,m-1,k}.
\end{eqnarray}
One also can write this in a symbolic manner introducing a kind of shifting operators with 
definition
\begin{eqnarray}\label{kmoperators}
\hat{k}{C'}_{nlmk}&=&{C'}_{nlm,k-1}, \cr
\hat{m}{C'}_{nlmk}&=&{C'}_{nl,m-1,k}.
\end{eqnarray}
With the use of Eq.(\ref{kmoperators}), Eq.(\ref{CviaZ}) may be rewritten as
\begin{eqnarray}
{C'}_{n+1,lmk}&=&(Z_{12}\hat{k}+Z_{11}\hat{m}){C'}_{nlmk}=
(Z_{12}\hat{k}+Z_{11}\hat{m})^2{C'}_{n-1,lmk}=\dots, \cr
{C'}_{n,l+1,mk}&=&(Z_{22}\hat{k}+Z_{21}\hat{m}){C'}_{nlmk}=
(Z_{22}\hat{k}+Z_{21}\hat{m})^2{C'}_{n,l-1,mk}=\dots,
\end{eqnarray}
and finally it can be written as
\begin{eqnarray}\label{finalsum}
{C'}_{nlmk}=(Z_{12}\hat{k}+Z_{11}\hat{m})^n (Z_{22}\hat{k}+Z_{21}\hat{m})^l {C'}_{00mk}=\cr
(\sum_{m'=0}^{n}\sum_{t'=0}^{l}C_{n}^{m'}C_{l}^{t'}Z_{11}^{m'} Z_{12}^{n-m'}
Z_{21}^{t'}Z_{22}^{l-t'}\hat{k}^{n+l-(m'+t')}\hat{m}^{m'+t'}){C'}_{00mk}.
\end{eqnarray}
One should note that, since total momentum of vacuum state should be zero, 
${C'}_{00mk}=0$
unless $m=k=0$. Therefore, in Eq.(\ref{finalsum}) only terms 
with ($m'+t'=m,n+l-(m'+t')=k$)
must survive and from Eq.(\ref{Zdef}) we get $Z_{11}=-Z_{22},
Z_{12}=Z_{21}$ to find:
\begin{equation}\label{appA01}
|0,\theta >={\cal Z}\sum_{n,l=0}^\infty \sum_{m=0}^{n+l}
\frac{B_{nlm}}{n!l!}(\hat a_1^{\dagger })^n (\hat a_2^{\dagger })^l
(\hat b_{-1}^{\dagger })^m(\hat b_{-2}^{\dagger })^{n+l-m}|0>,
\end{equation}
where
\begin{eqnarray}
B_{nlm}&=& \sum_{
\begin{array}{c}
m'+t'=m \cr 
0\leq m'\leq l \cr
0\leq t'\leq n
\end{array}}
C_n^{m'} C_l^{t'} Z_{11}^{l+m'-t'}Z_{12}^{n-m'+t'} (-1)^{l-t'}.
\end{eqnarray}
Using a direct expansion, one can also verify that the above expression for 
vacuum state is equivalent to
\begin{equation}\label{appA02}
|0,\theta >={\cal Z} \sum_{n,l}\frac{1}{n!l!}( Z_{11} 
a_1^{\dagger}b_{-1}^{\dagger } +Z_{12}a_1^{\dagger }b_{-2}^{\dagger })^n
(-Z_{11} a_2^{\dagger }b_{-2}^{\dagger }+Z_{12} a_2^{\dagger }b_{-1}^{\dagger })^l|0>. 
\end{equation}

\setcounter{equation}{0}
\setcounter{figure}{0}
\renewcommand{\theequation}{\mbox{B.\arabic{equation}}}
\renewcommand{\thefigure}{\mbox{B.\arabic{figure}}}
\section{Perturbative Expansion in $\theta$ for the flavor vacuum }
In this Appendix, we try to directly estimate the norm of the flavor 
vacuum state 
$G^{-1}\left(\theta \right) |0>=\exp \left( -\theta \hat S\right) |0>$ 
using the perturbative
expansion in powers of $\theta$ and show that the perturbative calculation
of the flavor vacuum state is indeed impossible for large $\gamma_{-} 
\theta$.
Truncating the series for $G^{-1}\left( \theta \right) |0>$ to the $N$ 
terms, we have the term with the largest number of particle coming from 
$\frac{\gamma _{-}}2\left( a_1^{\dagger }b_{-2}^{\dagger }+
a_1^{\dagger }b_{-2}^{\dagger}\right)$ 
in the $\left( -\theta \hat S\right) ^N$. Thus, the truncated series
of $G_N^{-1}\left( \theta \right) |0>$ can be written as
\begin{equation}
\begin{array}{c}
G_N^{-1}\left( \theta \right) |0>=X+\frac 1{N!}\left( -
\frac{\gamma _{-}\theta }2\right) ^N\left( a_1^{\dagger }b_{-2}^{\dagger
}+a_2^{\dagger }b_{-1}^{\dagger }\right) ^N|0> \\ =X+\frac 1{N!}\left( -%
\frac{\gamma _{-}\theta }2\right) ^N\sum_{n=0}^NC_N^n\left( a_1^{\dagger
}\right) ^n\left( b_{-2}^{\dagger }\right) ^n\left( a_2^{\dagger }\right)
^{N-n}\left( b_{-2}^{\dagger }\right) ^{N-n}|0>,
\end{array}
\end{equation}
where $X$ denotes all terms with the total number of particles and
antiparticles less than $2N$. For the norm of above expression we can write
then:
\begin{equation}
\begin{array}{c}
\left\| G_N^{-1}\left( \theta \right) |0>\right\| ^2=\left\| X\right\|
^2+\left( 
\frac{\gamma_{-}\theta }2\right) ^{2N}\frac 1{N!^2}\sum_{n=0}^Nn!n!\left(
N-n\right) !\left( N-n\right) ! \frac{N!^2}{n!^2\left(
N-n\right) !^2} \\ =\left\| X\right\| ^2+\left( N+1\right) \left( 
\frac{\gamma_{-}\theta }2\right) ^{2N}.
\end{array}
\end{equation}
Thus when $\gamma _{-}\theta >2$ the norm of the 
$|0,\theta>_N=G_N^{-1}\left( \theta \right) |0>$ is growing 
as the number of terms
kept in expansion of the $G\left( \theta \right)$ grows and therefore the
transformation operator $G_N^{-1}\left( \theta \right) $ is not well defined
operator in the mass-eigenstates Fock space.

One may also try to check directly the identity
$G\left( \theta \right) G^{-1}\left(\theta \right) =1$. 
In this type of approach one defines 
\begin{equation}
G\left( \theta \right) =\lim _{N\rightarrow \infty }G_N\left( \theta \right)
=\lim _{N\rightarrow \infty }\sum_{n=0}^N\frac{\left( \theta \cdot \hat
S\right) ^n}{n!}. 
\end{equation}
Then, one shall prove that 
$\lim_{N\rightarrow \infty }\left\| G_N\left(\theta \right) G_N^{-1}\left( \theta \right) -\hat 1\right\| =0$, 
{\it i.e.} $\lim_{N\rightarrow \infty }\left\| \left( G_N\left( \theta \right)
G_N^{-1}\left( \theta \right) -\hat 1\right) |x>\right\| =0$ for any
mass-eigenstate state $|x>$ if $G\left( \theta \right)$ is well defined.
When multiplying $G_N\left( \theta \right) $
and $G_N^{-1}\left( \theta \right) $ one typically get all coefficients
vanished till the power of $N$ and then have a ''tail'' up to the $\hat
S^{2N}$ coefficient. If $G\left( \theta \right)$ is well defined, this
tail is expected to vanish when $N$ is taken to infinity. However, 
this does not always happen in the perturbative expansion.
To demonstrate this one may consider the last term of the "tail" given
exactly by $\frac{\hat S^{2N}}{N!N!}$. Recalling that $\hat S$ generator
contains $\frac{\gamma _{-}}2\left( a_1^{\dagger }b_{-2}^{\dagger
}+a_2^{\dagger }b_{-1}^{\dagger }\right) $ combination, 
we can write the state $\left(G_NG_N^{-1}-1\right) |0>$ as 
\begin{equation}
\left( G_NG_N^{-1}-1\right) |0>=Y+\frac{\left( \theta \gamma _{-}\right)
^{2N}}{2^{2N}N!^2}\sum_{t=0}^{2N}C_{2N}^t\left( a_1^{\dagger }\right)
^t\left( b_{-2}^{\dagger }\right) ^t\left( a_2^{\dagger }\right)
^{2N-t}\left( b_{-1}^{\dagger }\right) ^{2N-t}|0>, 
\end{equation}
where $Y$ denotes all states with less then $4N$ number of particles
and antiparticles. The norm of this state is then given by
\begin{equation}
\begin{array}{c}
\left\| \left( G_NG_N^{-1}-1\right) |0>\right\| ^2=\left\| Y\right\|
^2+\frac 1{N!^4}\left( 
\frac{\gamma _{-}\theta }2\right) ^{2N}\sum_{t=0}^{2N}\left( C_{2N}^t\right)
^2 t!t!\left( 2N-t\right) !\left( 2N-t\right) ! \\ >\left(
2N+1\right) \left( \frac{\gamma _{-}\theta }2\right) ^{2N}. 
\end{array}
\end{equation}
Again when $\gamma _{-}\theta >2$ the above expression is not convergent and
the $G\left( \theta \right) G^{-1}\left( \theta \right)$ expression is in
fact not well defined in terms of mass-eigenstate fields.

For small values of $\theta$, however, the perturbative expansions are
indeed convergent and the radius of convergence is related to the pole
of ${\cal Z}=<0|0,\theta>$ on the complex plane of $\theta$. Here, the 
pole (critical) value is given by $\gamma_{-}\sinh(\theta_{critical})=2$.

\setcounter{equation}{0}
\setcounter{figure}{0}
\renewcommand{\theequation}{\mbox{C.\arabic{equation}}}
\renewcommand{\thefigure}{\mbox{C.\arabic{figure}}}

\section{Remarks on the Green function method and the arbitrary mass 
parametrization}

\subsection{Green Function Method}
We note that a straightforward use of the Green function
with the conventional definition 
$<0|T[\psi(x)\bar{\psi}(y)]|0>$
encounters some difficulties in the mixing analysis due to the fact that
the flavor vacuum state is not stationary in time 
($|0,\theta>(t)\neq|0,\theta>(t')$). The conventional Green function  
cannot be adopted without specifying at which times the flavor vacua were 
taken in the inner product. In fact, the most obvious generalization of 
the Green function 
as the overlap between the states created at times 
$x^0$ and $y^0$ 
({\it i.e.} $G(\alpha\rightarrow\beta;x^0,y^0) 
\\ =<0,y^0|T[\psi_{\beta}\bar{\psi}_{\alpha}]|0,x^0>$)
breaks down due to the unitary inequivalence of the
flavor Fock spaces at different times. Therefore,
some sort of modification, like parallel translation of states 
to the same time, shall be needed to define the Green function appropriate
in the mixing analysis. The flavor mixing problem can then be treated 
using this modified propagation functions as discussed in the previous 
literature\cite{8}.

In the process of our calculations we also noticed that some entities indeed
appeared as "transition" amplitudes from one state to another. For example, 
$<N_{\alpha}>=Z_{\alpha}+|A_{\alpha\alpha}|^2$
can be considered as superposition of "vacuum rotation" background contribution
$Z_{\alpha}$ and contribution from $\alpha\rightarrow\alpha$ transition
with $A_{\alpha\alpha}$ transition amplitude. In this manner, one can introduce the
Green function that only accounts for the transition amplitude
without the vacuum contribution. In this way Green function can be defined 
by:
\begin{equation}
G_{\alpha\alpha}(x,t;y,0)=<0'|\varphi_{\alpha}(x,t)\varphi_{\alpha}^{\dag}(y,0)|0'>
\end{equation}
where the vacuum state is taken at any (but certain) fixed time, for example 
$t=0$, and this coincides with the definition given in \cite{8}.
For the propagator with a definite momentum $\vec{k}$, we then obtain:
\begin{eqnarray}
G_{\alpha\alpha}(\vec{k},t)&=&A_{\alpha\alpha}^{*}(\vec{k},t),\cr
G_{\alpha\beta}(\vec{k},t)&=&A_{\beta\alpha}^{*}(\vec{k},t), etc.
\end{eqnarray}
Eq.(\ref{commtime}) then allows to define the propagation functions for 
any kind of transition. 

We should note however that the treatment with such a modified Green 
function does not cover all the variety of the effects in the mixing 
problem. In particular the condensate contribution $Z_{\alpha}$, that can be related to 
the unitary inequivalence of the flavor Fock spaces at different times, is lost
so that this part of problem is missing when the above approach is taken.
Nevertheless, the Green function method is useful in 
calculations of flavor-operator expectation values, scattering amplitudes 
and other quantities in which vacuum contribution $Z$ cancels out.

\subsection{Arbitrary Mass Parametrization}

In this sub-Appendix, we remark on the arbitrary mass 
parametrization\cite{Blasson3,ref3} in the mixing problem. 
As discussed in Ref.\cite{ref3}, one may treat
the flavor fields that were initially written as
\begin{equation}\label{arb01}
\phi _\alpha =\int \frac{d\vec k}{(2\pi )^{3/2}}
\left(u_{k,i}a_{k,i}(t)+v_{-k,i}b_{-k,i}^{\dagger }(t)\right) e^{i\vec k\vec x},
\end{equation}
equally well in an arbitrary mass basis,{\it i.e.}
\begin{equation}\label{arb02}
\phi_\alpha =\int \frac{d\vec k}{(2\pi )^{3/2}}\left( \tilde
u_{k,i}\tilde a_{k,i}(t)+\tilde v_{-k,i}\tilde b_{-k,i}^{\dagger }(t)\right)
e^{i\vec k\vec x},
\end{equation}
where $\tilde u_{k,i}$ and $\tilde v_{k,i}$ are free-field amplitudes with
some new arbitrary masses. Since there is no physical reason to prefer one
form over the other, Ref.\cite{ref3} and then 
Ref.\cite{Blasson3} claimed that no arbitrary mass parameters should 
appear in physically observable quantities,{\it i.e.} they shall be 
invariant under specific Bogoliubov transformation going from 
Eq.(\ref{arb01}) to Eq.(\ref{arb02}) \cite{Blasson3}
\begin{equation}\label{arb03}
\left( 
\begin{array}{c}
\tilde a_i(t) \\ 
\tilde b_i^{\dagger }(t)
\end{array}
\right) =J^{-1}(t)\left( 
\begin{array}{c}
a_i(t) \\ 
b_i^{\dagger }(t)
\end{array}
\right) J(t).
\end{equation}

It is true\cite{Blasson3,ref3} that the perturbative vacuum treatment 
\cite{6,7} yields the normalization of the flavor state not as unity but as 
some constant that depends on the arbitrary mass parameter. In 
particular, Eq.(\ref{ladden}) can be viewed as an
expansion of the flavor ladder operator in some basis constructed from the
free-field ladder operators. Then, the normalization of the one-particle 
state in perturbative vacuum treatment \cite{7} was given only by 
$|S_{-}|^2$ coefficient at the $b_{1,2}^{\dagger }$ operator, that is 
obviously depending on the choice of basis, e.g. changes with rotation in 
\{($a_1,a_2$)\&($b_1,b_2$)\} plane. Such arbitrariness is completely avoided 
in the exact vacuum treatment because the normalization of the flavor 
state is given by unity no matter what mass-basis is used. 

However, the claim\cite{Blasson3,ref3} that the number expectation 
values are not physical because they do depend on the arbitrary mass 
parameters cannot be correct. As a counter example to such claim, one can 
consider a very specific case of the mixing problem, namely 
when the mixing is absent ($G(\theta )=1$). As we discuss below,
applying such claim\cite{Blasson3,ref3} to this specific example 
leads to a conclusion that cannot be correct. 
With no mixing, we are dealing with nothing else but free-field problem 
where the particle number operator 
$N_i(t)=a_i^{\dagger }(t)a_i(t)$ and the particle number expectation value 
$<N_i>=|\{a_i(0),a_i^{\dagger }(t)\}|^2$ (for simplicity we
consider fermion case) must be well-defined physically 
observable quantities. However, when we apply the transformation given by 
Eq.(\ref{arb03}) and compare directly Eq.(\ref{arb01}) and 
Eq.(\ref{arb02}), we observe that 
\begin{equation}
\tilde a_i=(\tilde u_{k,i}^{\dagger }u_{k,i})a_{k,i}e^{-i\epsilon
_{k,i}t}+(\tilde u_{k,i}^{\dagger }v_{-k,i})b_{-k,i}^{\dagger }e^{i\epsilon
_{k,i}t}=\rho _ka_{k,i}e^{-i\epsilon _{k,i}t}+\lambda _kb_{-k,i}^{\dagger
}e^{i\epsilon _{k,i}t} 
\end{equation}
and
\begin{equation}
<\tilde N_i>=|\{\tilde a_i,\tilde a_i^{\dagger }(t)\}|^2=||\rho
_k|^2e^{-i\epsilon _{k,i}t}+|\lambda _k|^2e^{i\epsilon _{k,i}t}|^2,
\end{equation}
where we follow the notations in Refs.\cite{Blasson3,ref3}.
In the case of free fields the number expectation value does 
depend on the arbitrary mass parameters and thus following the above 
claim one may conclude that $<N_i>$ is not a physically measurable 
quantity. However, this cannot be correct because both the particle 
number operator and the number expectation value in the free-field problem
are well-defined physical observables.
We viewed the above inconsistency as follows.
The transformation given by Eq.(\ref{arb03}) is in fact nothing else but
redefinition of the particle states, so that the tilde quantities 
correspond to some new quasiparticle objects and the number operator now 
describes the number of different type of particles than before. 
Therefore, the number operator average shall not be expected to be same  
in such transformations. Indeed, it should change in some covariant and  
self-consistent manner instead of being invariant under such a 
redefinition.

\end{appendix}

\end{document}